\documentclass[10pt,letterpaper]{article}
\usepackage[top=0.85in,left=2.75in,footskip=0.75in]{geometry}

\usepackage{changepage}

\usepackage[utf8]{inputenc}

\usepackage{textcomp,marvosym}

\usepackage{fixltx2e}

\usepackage{amsmath,amssymb}

\usepackage{cite}

\usepackage{microtype}
\DisableLigatures[f]{encoding = *, family = * }

\usepackage{rotating}

\usepackage{setspace} 
\raggedright
\usepackage[aboveskip=1pt,labelfont=bf,labelsep=period,justification=raggedright,singlelinecheck=off]{caption}

\makeatletter
\renewcommand{\@biblabel}[1]{\quad#1.}
\makeatother

\date{}

\usepackage{lastpage,fancyhdr,graphicx}
\fancyheadoffset[L]{2.25in}
\fancyfootoffset[L]{2.25in}
\begin{document}
\vspace*{0.35in}

\begin{flushleft}
{\Large
\textbf\newline{Evidence for a creative dilemma posed by\\ repeated collaborations}
}
\newline
\\
Hiroyasu Inoue\textsuperscript{1,*}
\\
\bigskip
\bf{1} Graduate School of Simulation Studies, University of Hyogo, Kobe, Hyogo, Japan.
\\
\bigskip

%
%





* inoue@sim.u-hyogo.ac.jp

\end{flushleft}
\section*{Abstract}
We focused on how repeat collaborations in projects for inventions
affect performance.
Repeat collaborations have two contradictory aspects.
A positive aspect is team development or experience,
and a negative aspect is team degeneration or decline.
Since both contradicting phenomena are observed,
inventors have a dilemma as to whether they should keep collaborating in a team or not.
The dilemma has not previously been quantitatively analyzed.

We provide quantitative and extensive analyses of the dilemma in creative projects
by using patent data from Japan and the United States.
We confirm three predictions to quantitatively validate the existence of the dilemma.
The first prediction is that the greater the patent a team achieves, the longer the team will work together.
The second prediction is
that the impact of consecutive patents decreases after a team makes a remarkable invention, which is measured by the impact of patents.
The third prediction is that the expectation of impact with new teams is greater than that with the same teams successful in the past.
We find these predictions are validated in patents published in Japan and the United States.
On the basis of these three predictions, we can quantitatively validate the dilemma in creative projects.
We also propose preventive strategies for degeneration.
One is developing technological diversity, and another is developing inventor diversity in teams.
We find the two strategies are both effective by validating with the data.



\section*{Introduction}

Our globally connected societies require us to be
aware of competitiveness in various levels, such as individuals, companies, or countries \cite{Porter98}.
Knowledge creation has attracted great interest as a way to be competitive
instead of incorporating labor-intensive processes that are typically observed in manufacturing \cite{Grant96, Hall01, McEvily02}.
Since the central resource of knowledge creation is knowledge workers \cite{Drucker92},
it is important to know how to support their activities.
Although it is recognized that a genius could
make great creations \cite{Bowler05},
recent studies have indicated teams can generate better outcomes than solo researchers on average \cite{Wuchty07}.
Recent studies have also revealed a rising propensity for teams rather than solo researchers \cite{Merton79, Jones05, Wuchty07}.
If teams can perform better,
our next question is what qualities in teams affect performance.


Repeat collaborations seem to be a key to understanding
how well teams work in creative projects,
and there is a growing field of study on repeat collaborations that occur in creative projects.
One reason for this interest is that
repeat collaborations can be understood as a process of coordination between creators.
For example, Skilton and Dooley indicated that
there is a sequence of processes constituted by idea generation,
disclosure/advocacy, and convergence, which they call ``creative abrasion'' \cite{Skilton10}.

Through a survey of the studies on repeat collaborations,
it seems the repeat collaborations have two contradictory aspects.
The first is positive: team development or experience.
Studies of team development using various models have described the kinds of processes
that allow team members to cultivate mutual relationships and improved performance \cite{Kozlowski03, Schwab08}.
These studies described in what kinds of processes team members
cultivated mutual relationships and improved performance.
The second is negative: team degeneration or decline.
Previous studies have found that repeat collaborations underperform in comparison with initial collaborations in creative projects,
e.g., scientific research \cite{Porac04,Guimera05,Inoue15}, 
consulting practice \cite{Reagans04}, and performances in entertainment \cite{Guimera05, Delmestri05, Uzzi05, Perretti07}.

Repeat collaborations are particularly discussed in psychology.
The terms ``habitual routines'' or ``behaviour of groups'' are used to mean repeat collaborations.
The first review for habitual routines of groups seems to be given by Gersick and Hackman \cite{Gersick90},
though they said the paper did not provide either a literature review or a theory of habitual behavior
but rather presented a broad-brush survey.
The survey pointed out that
``habitual routines can reduce the likelihood of innovative performance processes.''



This contradiction is not just a theoretical conflict.
The contradiction is descriptively discussed
in regard to innovation in companies in  ``The Innovator's Dilemma'' \cite{Christensen97}.
This book, which contains numerous observed examples, explains
how new companies with disruptive technologies
redefine competitiveness in markets,
and successful and preexisting companies cannot adjust themselves to changes because of past successes.
Companies with successful products are fixated on their successes and end up ruined in the end.
Although what we discuss is not the dilemma of companies but rather individuals,
they share the same basis.

In this paper, we investigate
extensive and quantitative analyses of repeat collaborations on patent applications
by using data of Japan and the United States.
In particular,
we define the dilemma as the phenomenon
that a successful team is fixated on repeat collaborations even if performance declines
and members of the team lose the chance for greater success in different teams.
There have not been quantitative analyses of this dilemma with creative projects.



The psychological studies, at their core, try to reveal the mechanisms behind repeat collaborations.
To the contrary, we intentionally avoid discussing the mechanisms
because we do not have detailed data such as on communication in teams. 
Also, the difference between the psychological studies and our study can be attributed to data.
Their data is detailed but basically lab-scale.
In comparison, our data is longitudinal and wide-scope.
The difference means the results complement each other.

It is in the nature of teams that they have routines because routines
enable us to exploit the knowledge and coordination within teams
and avoid unnecessary costs in rebuilding this knowledge or coordination.
Therefore, how we can avoid the gradual failure of performance during routines is a practical issue.
Obviously, some stimuli are necessary to avoid gradual failure.
We consider introducing new technological fields into teams and mixing team memberships as the stimuli
and investigate the effect of the stimuli by our data.
In particular, mixing team memberships has already been studied in the psychology field \cite{Gorman10, Gorman11}.

The paper is organized as follows.
First, we describe the data we use.
Second, we describe how we analyze the dilemma with patent data from Japan and the United States and
quantitatively demonstrate the dilemma of inventor teams.
Third, we propose strategies to prevent degeneration and explain how we validate the strategies on the basis of data.
Finally, we conclude the paper.

\section*{Data}

Patent data is suitable
for studying creative projects.
This is because patents indicate the occurrence of innovations over time \cite{Griliches98}.
Also, they contain massive data on repeated collaborations.
We use the Institute of Intellectual Property (IIP) patent database to obtain the Japanese patent data \cite{Goto07},
and the National Bureau of Economic Research (NBER) U.S. patent citations data and Patent Network Dataverse
to obtain the United States patent data \cite{Hall01, Lai12}.
There is other patent data published by the European Patent Office, which is called ``PATSTAT''.
We do not use PATSTAT because inventors are not identified and much work is required to deal with it.
Our approach using massive data to help us analyze societies as complex systems,
is in line with the ``computational social science'' framework \cite{Lazer09}.

We can extract common data
so that we can compare the two databases.
The data have the IDs of inventors
who applied for patents,
the number of received citations,
the technological classifications of patents,
and the year of application.
We use the International Patent Class for Japanese patent data
and the U.S. patent class for U.S. patent data to get the technological classifications.
To quantify the performance of patents,
we define the impact ($I$) of a patent
by using the number of citations \cite{Hall01}.
Since older patents have more chances of being cited,
impact is the number of citations divided by the average number of citations
of patents granted in the same year.

Here, teams are defined as assemblies of more than one individual.
Even if a team is a subset of another team, that team is considered to be a different team.
For example, if there is a sequence of patents applied for for patent 1 by inventors A and B
and patent 2 by inventors A, B, and C, we count two distinct first-patents as those of team A and B and team A, B, and C.
Though team A and B is a subset of team A, B, and C, those teams are considered as different teams.
One may argue that the collaboration between inventors A and B is not the first time at patent 2,
and therefore, it is not appropriate to count them as in a different team.
This viewpoint may be correct.
On the other hand, it is doubtful that these two collaborations (without inventor C and with inventor C) are the same team.
Therefore, we here count a collaboration when a setup of inventors is completely the same.
In a later section, we will revisit this issue and redefine how we count teams.

Table \ref{tbl:overview} summarizes
fundamental data.
The data show that there are enough patents and citations between patents
to statistically discuss the topics.
Regarding the time frame of the data,
we use the maximum time frame provided by the databases because longitudinal observation is necessary for our study.

\begin{figure*}[h!]
 \begin{center}
    \includegraphics[scale=0.6, bb=0 0 500 500]{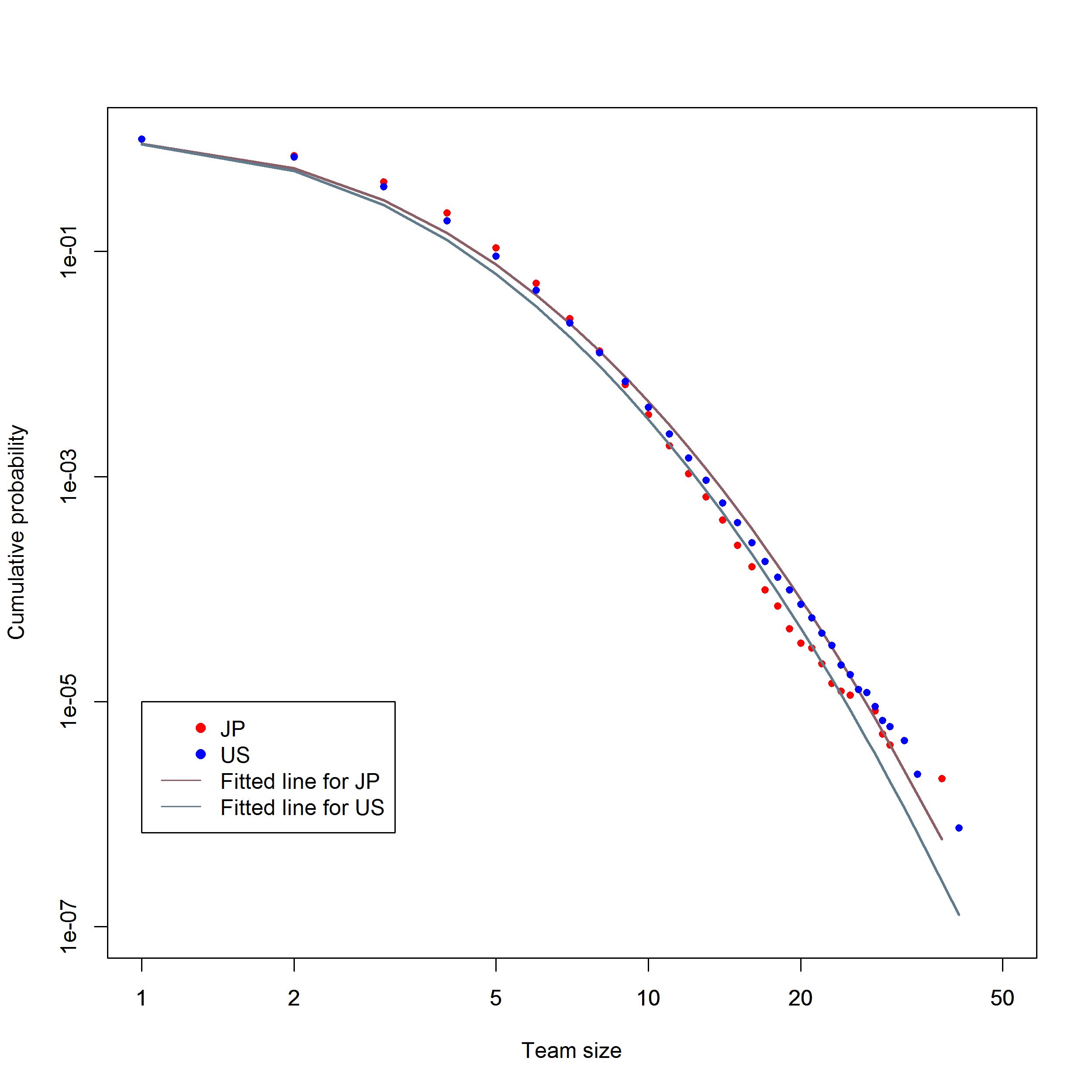}
  \end{center}
  \caption{Cumulative probability distribution of team size. Horizontal axis shows team size. Vertical axis shows cumulative probability. Lines are fitted to log-normal distributions.\label{fig:teamsize}}
\end{figure*}

Since we mainly discuss the impact of team inventions,
it is worth checking their distributions.
Fig. \ref{fig:teamsize} plots
the cumulative probability distributions of team size on the data.
The vertical axis shows cumulative probability.
The horizontal axis shows team size.
The red plots are for Japanese data, and the blue ones are for the U.S. data.
The lines are fitted to log-normal distributions by using maximum likelihood estimation.
The grayish red line is for Japan, and the grayish blue line is for the U.S.
The lines seem to fit well to the plots.

\begin{figure*}[h!]
 \begin{center}
    \includegraphics[scale=0.6, bb=0 0 500 500]{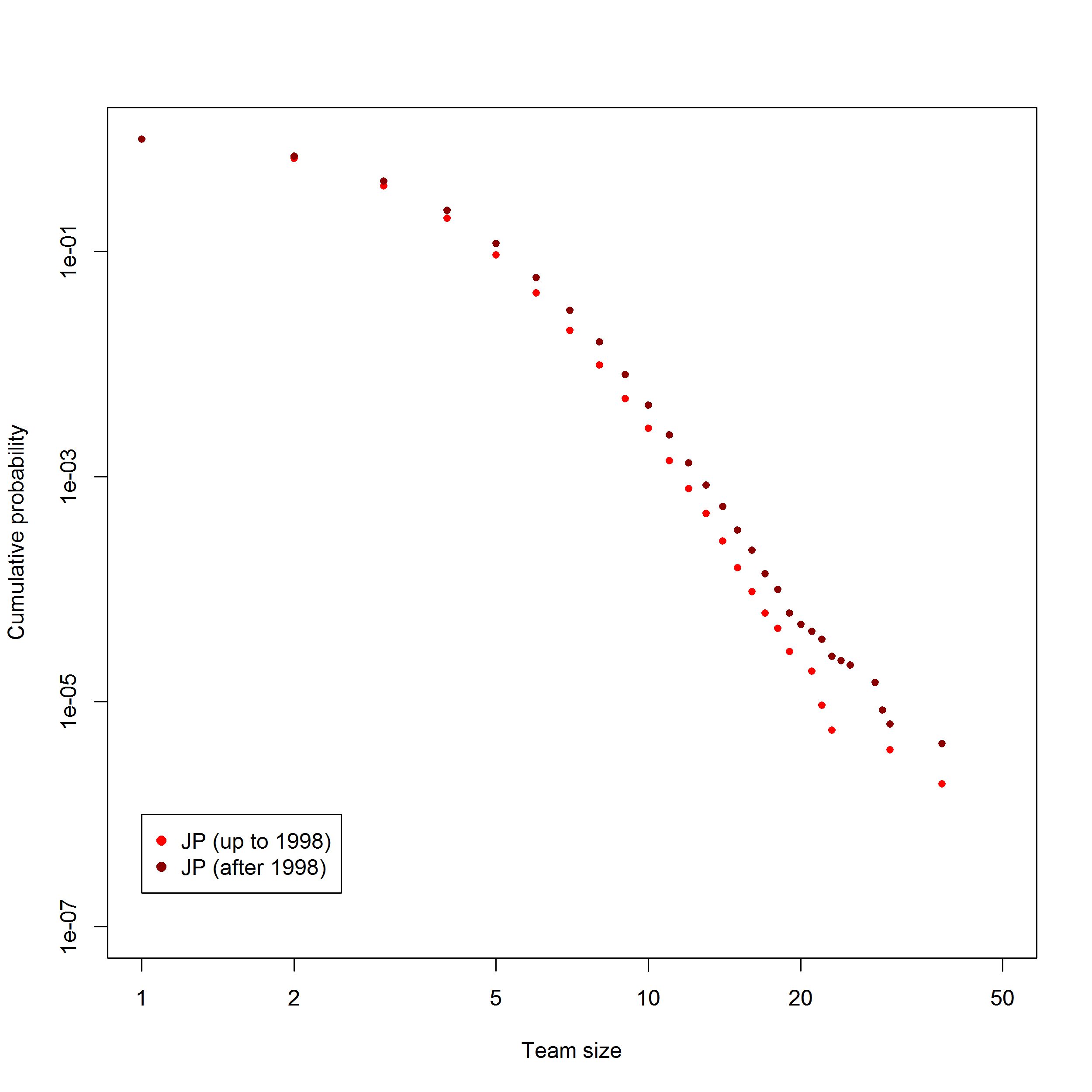}
  \end{center}
  \caption{Cumulative probability distribution of team size in Japan. Horizontal axis shows team size. Vertical axis shows cumulative probability. Data is separated around middle of duration. \label{fig:teamsizeSepJp}}
\end{figure*}

\begin{figure*}[h!]
 \begin{center}
    \includegraphics[scale=0.6, bb=0 0 500 500]{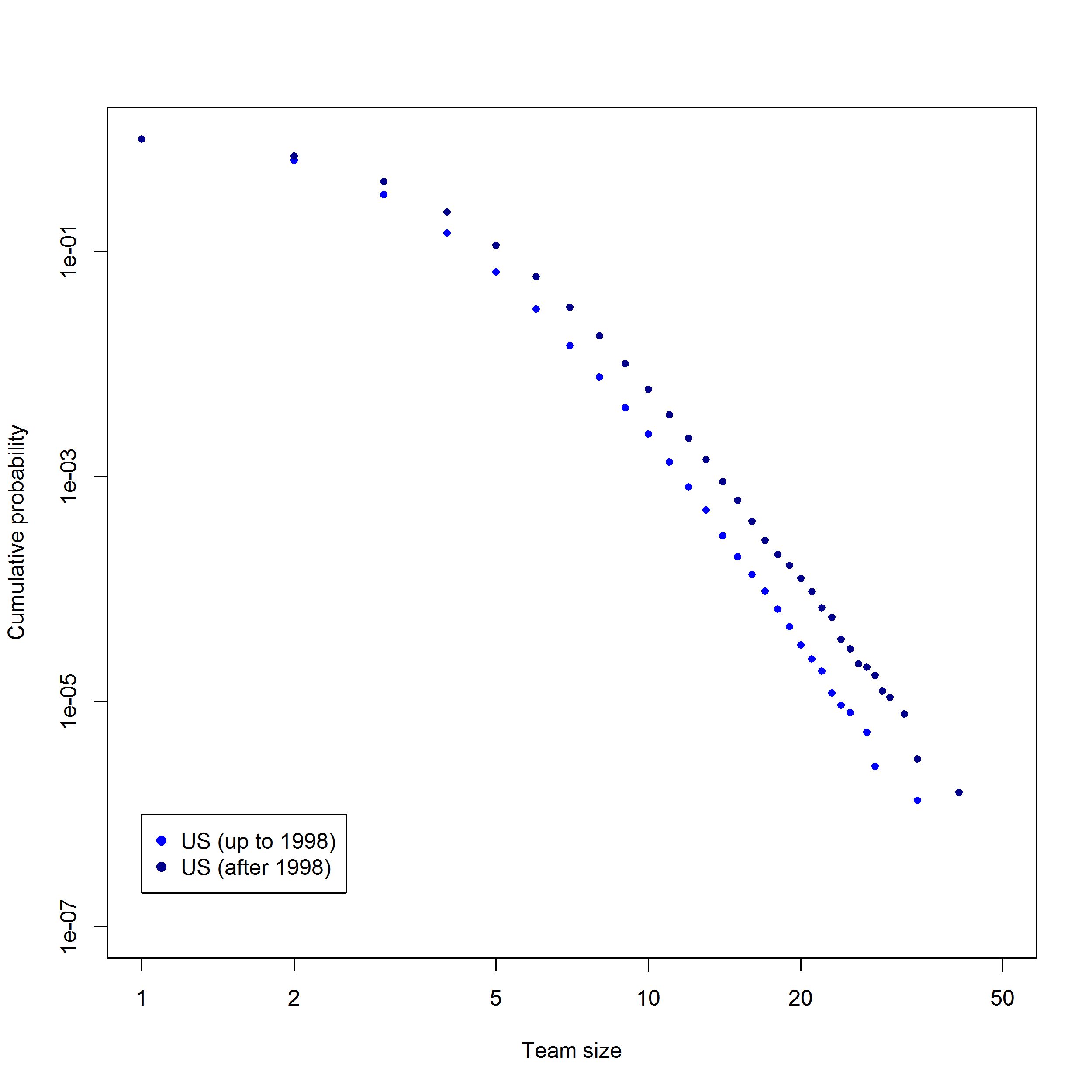}
  \end{center}
  \caption{Cumulative probability distribution of team size in the U.S. Horizontal axis shows team size. Vertical axis shows cumulative probability. Data is separated around middle of duration. \label{fig:teamsizeSepUs}}
\end{figure*}

Fig. \ref{fig:teamsizeSepJp} and
Fig. \ref{fig:teamsizeSepUs} show the cumulative distributions of team size
in the first and second halves of the covered time periods, for the Japan and U.S. data, respectively.
The meanings of figures are the same as Fig. \ref{fig:teamsize}.
As we can see, team sizes are growing in both countries.
The result is in line with the indications given by preceding studies \cite{Guimera05, Wuchty07}.

\begin{figure*}[h!]
\begin{center}
  \includegraphics[scale=0.6, bb=0 0 500 500]{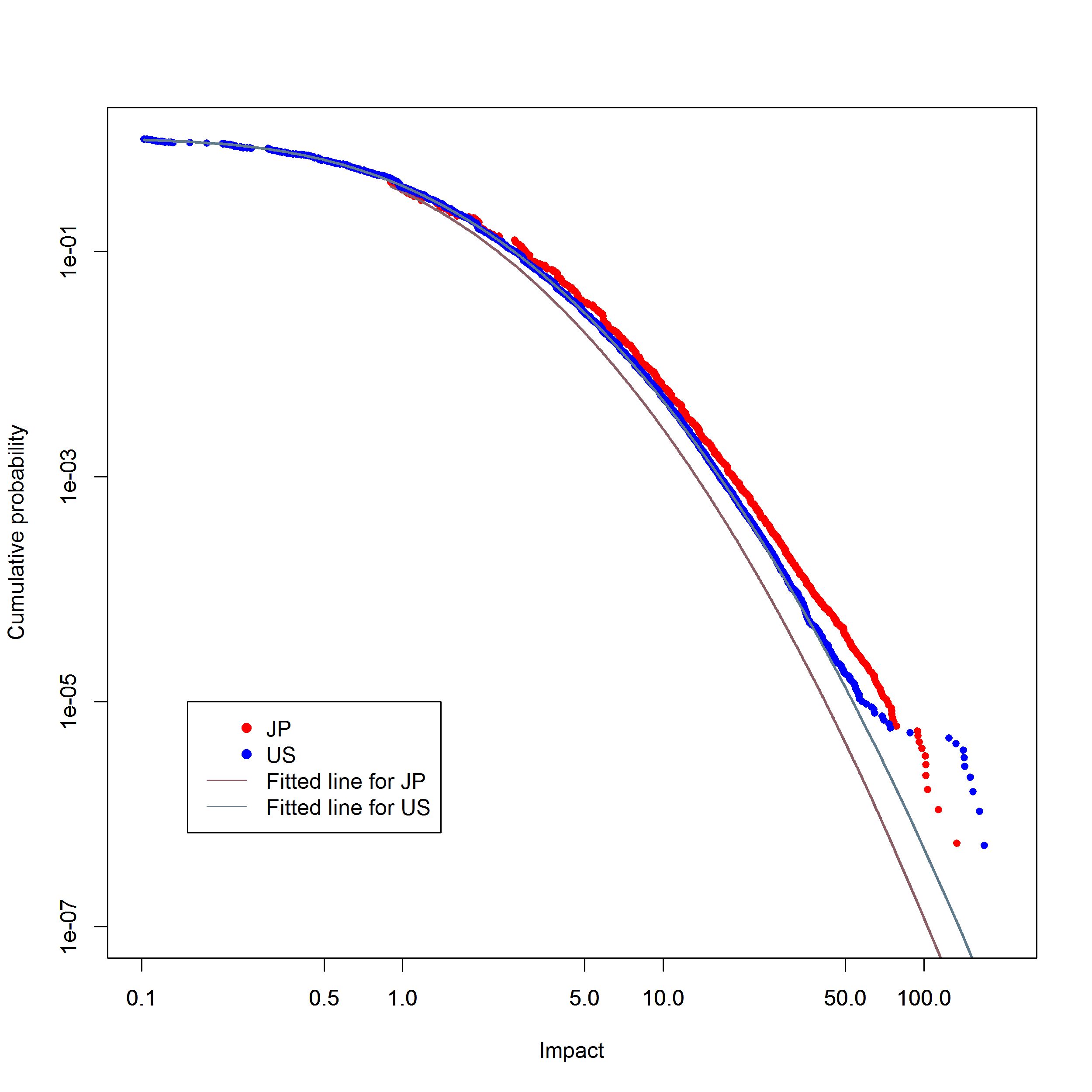}
\end{center}
\caption{Cumulative probability distribution of impact. Horizontal axis shows impact. Vertical axis shows cumulative probability. Lines are fitted to log-normal distributions. \label{fig:impact}}
\end{figure*}

Fig. \ref{fig:impact} plots
the impact distributions for the datasets.
The meanings of the plots and fitted lines are the same as those in Fig. \ref{fig:teamsize}.
It is interesting that 
there are no large deviations between datasets
even if they are covered by different patent laws in their countries.
Regarding fitted lines,
the U.S.'s distribution is well fitted by a log-normal distribution.
In comparison, the Japanese distribution is not fitted as well as the U.S. one.
Since the impact less than one is not frequently seen in Japan,
it seems that the absence causes the deviation of the fitted line for Japan.
Detailed investigation into the cause is beyond this paper,
but we just add that this difference can happen since citation dynamics depend on the cultures of countries.

\begin{table*}[h!]
\caption{Overview of datasets. Datasets used are Japanese (JP) and
United States patent data (US).
Range of years in which patents were applied for is labeled ``duration.''
In addition, table lists numbers of patents, inventors, teams, and citations.}
\begin{center}
\label{tbl:overview}
\begin{tabular}{l|r|r|}
Datasets & \multicolumn{1}{l|}{JP} & \multicolumn{1}{l|}{US} \\
\hline
Duration (year) & \multicolumn{1}{l|}{1964-2012} & \multicolumn{1}{l|}{1975-2010} \\
No. of patents & 4,349,161 & 3,984,771 \\
No. of inventors & 1,538,525 & 2,665,7091 \\
No. of teams & 967,159 & 1,325,869 \\
No. of citations & 18,410,996 & 48,911,485 \\
\end{tabular}
\end{center}
\end{table*}

\section*{Quantitative demonstration of the dilemma}

\label{cha:dilemma}

One of the goals in this paper, again, is
to provide extensive and quantitative analyses of the dilemma with creative activities.
We explain how we find the dilemma in regard to creative activities.

A typical description of the dilemma is
that a successful team tends to try to reproduce the successes
and members of the team, therefore losing chances to achieve other successes in different teams.
We propose three predictions
to validate the dilemma.
\begin{itemize}
\item Prediction 1: The greater the patent a team achieves, the longer the team will work together.
\item Prediction 2: The impact of consecutive patents decreases after a team makes great patents (hits).
\item Prediction 3: At some point during consecutive patents,
the expectation of impact by switching teams is greater than that from consecutively working in the same team.
\end{itemize}
Prediction 1 means that a team's great patent (hit) tends to bind members for a long time.
Teams try to reproduce patents if they succeed, and vice versa.
Prediction 2 has already been reported in previous studies \cite{Guimera05,Inoue15}.
The opposite of Prediction 2 is that there is no decrease or, instead, an increase in impact after hits.
Prediction 3 means that
individuals in teams with past successes miss chances to produce better patents by switching teams.
If all of these three predictions are supported by the data, we can say that the dilemma occurs.

\begin{figure*}[h!]
  \begin{center}
    \includegraphics[scale=0.6, bb=0 0 500 500]{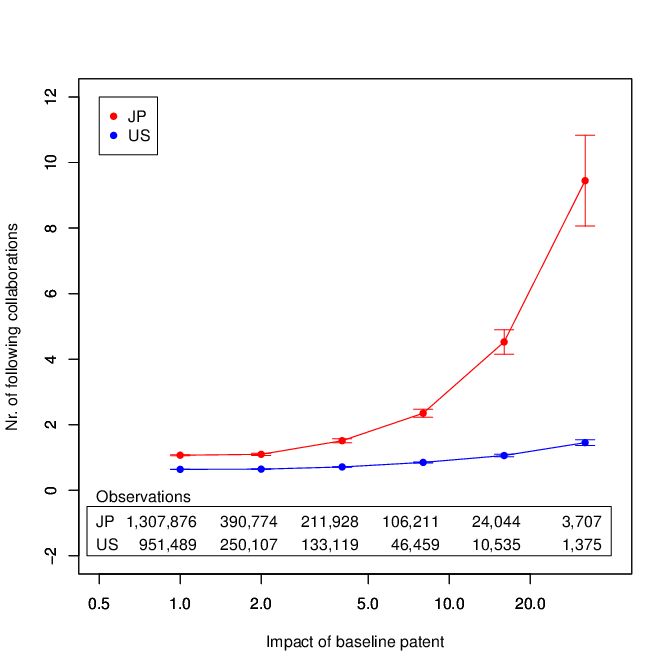}
  \end{center}
  \caption{Average number of patents made by a team after it achieves a patent of a given impact.
Horizontal axis is the impact of baseline patent.
Vertical axis is average number of subsequent patents by same team.
Brackets show standard errors.
Horizontal axis is cut so that number of samples is greater than approximately 100.
\label{fig:restRep}}
\end{figure*}

Fig. \ref{fig:restRep} plots
the average number of patents made by a team after it creates a patent of a given impact.
The impact of the first patent (the baseline patent) is shown on the horizontal axis,
while the vertical axis shows the average number of subsequent patents by the same team.
This average includes only the subsequent patents,
not the baseline patent.
Note that every patent by a team is used once as a baseline, and the numbers of subsequent patents for all baseline patents are counted.

We conduct linear regression analyses
to validate whether or not it is true that
the greater the impact of a team's patent, the more subsequent patents that the team will produce.
Though the relationships in Fig. \ref{fig:restRep} do not look linear,
as long as we have significantly positive results with the linear regression analyses,
the results satisfy what we need.
We find both data have positive and significant coefficients for impact
(Japan: coefficient $=0.0042$, $p<10^{-16}$;
U.S.: coefficient $=0.032$, $p<10^{-16}$).
Therefore, the better the impact is, the more patents the teams will have.
The results satisfy Prediction 1.

If the curves in Fig. \ref{fig:restRep} are concave,
the probability of repeating a collaboration increases as the impact of a baseline patent increases.
In addition, the more concave the curve is, the more sensitive the response is.
Therefore, teams for Japanese patents tend to have more subsequent patents, and the number is more sensitive to prior impacts
than that for the United States patents.
To statistically test whether Japanese inventors have more subsequent patents than U.S. inventors,
we conduct Wilcoxon rank-sum tests.
The reason we choose the Wilcoxon rank-sum test is that
the distributions of subsequent patents do not look like a normal distribution.
The test is one-sided.
For every separate bin of impact in Fig. \ref{fig:restRep},
we find subsequent Japanese patents are dominant at the five percent level.

\begin{figure*}[h!]
\begin{center}
  \includegraphics[scale=0.6, bb=0 0 500 500]{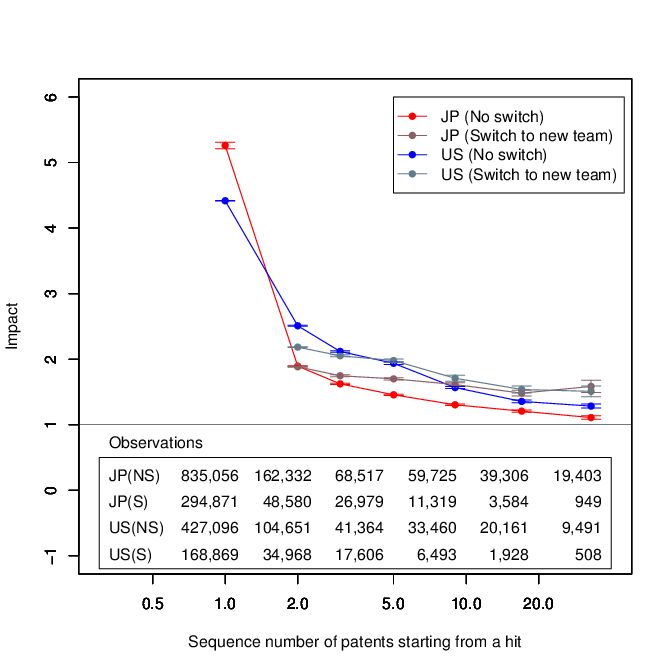}
\end{center}
\caption{Average impact of patents following hits.
We define hits as patents with top 10\% impact.
Horizontal axis is number of patents including a hit and all subsequent patents by the same team that made the hit.
The hit is not necessarily the team's first patent.
Vertical axis is average impact.
Horizontal axis is cut so that number of samples is greater than approximately 100. 
Red, grayish red, blue, and grayish blue lines correspond to Japanese ``no switch,''
 Japanese ``switch to new team,'' U.S.'s ``no switch,'' and U.S.'s ``switch to new team.''
Brackets show standard errors.
Black horizontal line is impact $=1$.
\label{fig:top10}}
\end{figure*}

Fig. \ref{fig:top10} plots the
average impact of subsequent patents after hits.
Here, we define hits as patents with top 10\% impact.
The vertical axis is the average impact. 
The horizontal axis shows a patent's place within the sequence that includes the hit and subsequent patents.
The red line is for Japan, and the blue line is for the U.S.
Those lines show repetitions by teams whose membership did not change (``no switch'').
As an overview of these results,
we can see that teams that made hits in the past tend to consecutively create beneficial patents
because those lines in Fig. \ref{fig:top10} are above one, the average impact.
Since we can see a fall in repetitions without switching teams,
the data fulfill Prediction 2.

The grayish lines in Fig. \ref{fig:top10} show
the average impact of the first patents applied for by a new team.
The grayish red line is for Japan, and the grayish blue line is for the U.S. (``switch to new team'').
The new team includes individuals who
belonged to teams that made hits.
Since only the first patents with new teams are considered,
repetitions for ``switch to new team'' given by the horizontal axis have a different meaning from ``no switch.''
Here, we want to compare the difference between ``no switch'' and ``switch to new team.''
If a member belonged to a team (with a hit) until repetition $r-1$
and then switched to a new team, the impact of the first patent of the new team is plotted at $r$.
By doing so,
we can compare the average impact of the repetition $r$ of ``no switch'' with the average impact of
the first patents of the team to which the member switched.


Comparing the lines between no switching and switching enables us to understand
when individuals should switch teams.
We conduct the Wilcoxon rank-sum tests between no switching and switching
to see when the average impacts of switching are significantly dominant over no switching.
As mentioned earlier, the reason we choose the Wilcoxon rank-sum test is that
the impact distribution is highly skewed.
The test is one-sided.
We found that switching is statistically dominant at and after the third patents at the five percent level in the Japanese data.
The tipping point for the U.S. is the fifth patent.
It should be noted that since samples sizes are not large enough after a lot of repetitions,
we cannot stably see the significance after the 11th patent for Japan and the 12th patent for the U.S.
As a result,
individuals should seemingly switch teams after the second patent
following a hit in teams for Japanese patents and after the fourth patent for U.S. patents.
The data satisfy Prediction 3 based on this discussion.

\begin{figure*}[h!]
  \begin{center}
    \includegraphics[scale=0.6, bb=0 0 500 500]{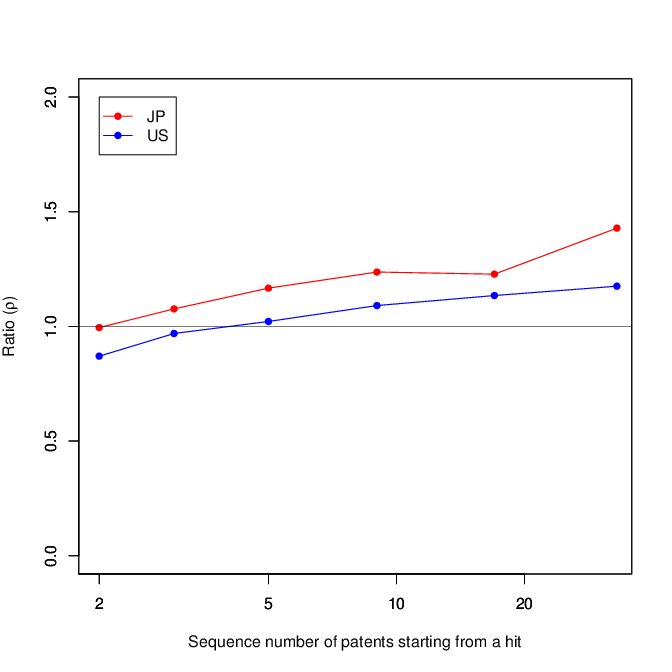}
  \end{center}
  \caption{Relationship between $\rho_{\mbox{r}}$ and repetition.
Horizontal axis is
number of patents including a hit and all subsequent patents by the same team
that made the hit.
Vertical axis is ratio ($\rho_{\mbox{r}}$).
Hits are patents with top 10\% impact.
Horizontal axis is cut so that number of samples is greater than approximately 100.
Black horizontal line is $\rho_{\mbox{r}}=1$.
\label{fig:rho10}}
\end{figure*}

To clearly see the effect of switching and discuss Prediction 3,
let us define a measure, $\rho_{\mbox{r}}$.\\
{\small
$\rho_{\mbox{r}} \equiv \frac{\mbox{Average impact of first patents created in new teams after (r-1)-th patents}}{\mbox{Average impact of r-th patents in teams}}$
}.\\
If $\rho_{\mbox{r}}$ is greater than one, the expectation of impact with the new teams is larger than the r-th patents in current teams.
Fig. \ref{fig:rho10} plots the
$\rho_{\mbox{r}}$ for each dataset.
Note that the plots are teams that had a hit, which is the same as Fig. \ref{fig:top10}.
The horizontal axis means r-th patents after hits in teams.
The definition of the repetition number is consistent with Fig. \ref{fig:top10}.
The vertical axis means $\rho_{\mbox{r}}$.
The black horizontal line is $\rho_{\mbox{r}}=1$.
As subsequent patents continue to be filed,
$\rho_{\mbox{r}}$s increase monotonically.
As was already pointed out,
$\rho_{\mbox{r}}$ seems to surpass 1 at the third patent of teams (the second patent after hits) for Japanese patents.
The transit point for U.S. patents is five (the fourth patent after hits).

\begin{figure*}[h!]
  \begin{center}
    \includegraphics[scale=0.6, bb=0 0 500 500]{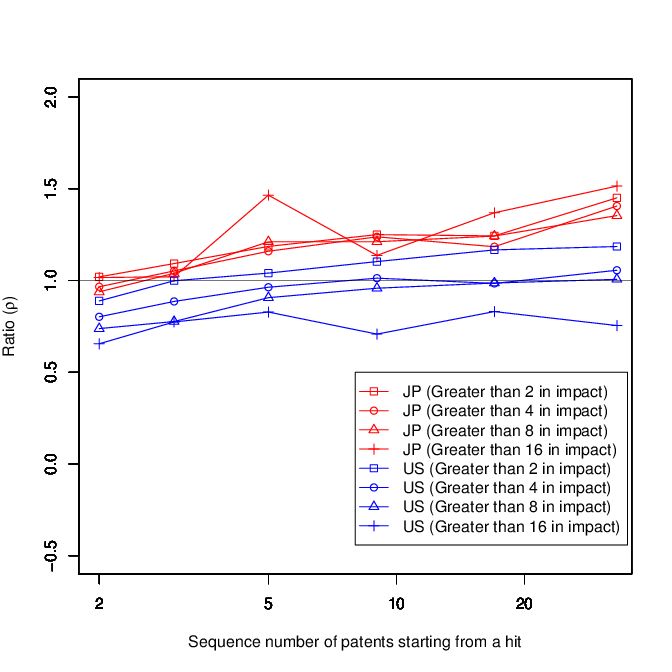}
  \end{center}
  \caption{
Effect on $\rho_{\mbox{r}}$ of hit threshold.
Horizontal axis is number of patents including a hit
and all subsequent patents by the same team that made the hit.
Vertical axis is ratio ($\rho_{\mbox{r}}$).
There are four different thresholds of hits in this figure: greater than 2, 4, 8, and 16 in impact. Black horizontal line is $\rho=1$.
\label{fig:sepRho}}
\end{figure*}

It is natural that the line of $\rho_{\mbox{r}}$ is dependent on the threshold of a hit.
We have set the threshold at top 10\% so far for simplicity.
We show lines of $\rho_{\mbox{r}}$ with different thresholds
in Fig. \ref{fig:sepRho}.
There are four lines, each with a different threshold, shown for both Japan and the U.S.
These thresholds are ``greater than 2,'' ``greater than 4,'' ``greater than 8,'' and ``greater than 16'' in impact
(note that ``greater than 2'' is roughly equivalent with the threshold for the top 10\%).
As is indicated in Fig. \ref{fig:sepRho}, we do not find significant difference between thresholds in Japanese data.
However, it seems that the higher the threshold of hit is, the lower the $\rho_{\mbox{r}}$ is in the U.S. data.
We conduct the Wilcoxon rank-sum tests between ``greater than 2'' and ``greater than 16'' for every repetition as a typical example.
We find that $\rho_{\mbox{r}}$ for ``greater than 16'' is significantly lower than that for ``greater than 2.''
We can say it is reasonable in the U.S. for people to postpone moving into a new team on the basis of the extent of impact.

Since all predictions were confirmed,
we can say we have quantitatively demonstrated the dilemma of inventors,
which has not been done before.

We have not discussed other possible variables that can be used instead of repetition thus far.
Possible variables include the age of inventors, the number of team patents, and the number of technology classes that a team took.
All these variables are related to team experience.
We already investigated these variables and found that repetition and age are significantly related to the impact of patents \cite{Inoue15}.
Although repetition and age are also correlated, we also found that each variable separately affects the impact.
Therefore, we can say repetition is a potent determinant of the impact of team patents.

\section*{Prevention of degeneration}

In the previous section, we explained that the creativity of teams declines on average,
and therefore, there is a dilemma for inventors as to whether to repeat or to switch teams.
Since switching teams involves the cost of communication needed to build a new relationship
and runs the risk of project failure,
alleviating degeneration in repeat collaborations can help inventors.
Here, we propose two strategies to do so.

\begin{figure*}[h!]
  \begin{center}
    \includegraphics[scale=0.6, bb=0 0 500 500]{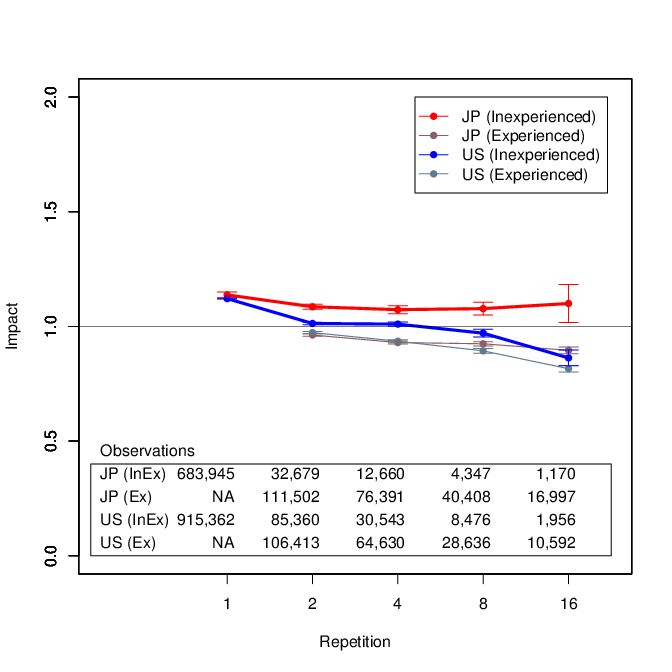}
  \end{center}
  \caption{
Average impact of repeat collaborations categorized by technological development.
Horizontal axis is repetitions of team patents.
Vertical axis is average impact.
Brackets show standard errors.
``Inexperienced with technology'' (InEx) means patents using technologies that teams have no experience with.
``Experienced with technology'' (Ex) means the opposite.
Horizontal axis is cut so that number of samples is greater than approximately 100.
Black horizontal line is impact $=1$. \label{fig:tech}}
\end{figure*}

The first strategy is developing technological diversity.
Fig. \ref{fig:tech} plots results with patents separated into two groups: inexperienced and experienced with technologies.
If a patent at some repetition number has a technology that a team has no experience with,
it is categorized into ``inexperienced with technology.''
Otherwise it is categorized into ``experienced with technology.''
The repetition number means the number of patents applied for by the same teams.
The plots include all teams regardless of whether patents are hits or not,
though we have discussed teams with hits thus far.
Since the results we show here are not limited to teams with hits,
the prevention strategies can be broadly applied to all teams.
The results reveal that repetitions in the inexperienced category are better than those in the experienced category.
Moreover, decline itself does not occur in Japanese patent data.

We conduct the Wilcoxon rank-sum tests between inexperienced and experienced with technology.
We find that inexperienced is statistically dominant at the five percent level at every repetition in both the Japanese and U.S. data.
Note that since there are not large enough sample sizes after a lot of repetitions,
we cannot stably see the significance after the ninth patent for Japan and the 33rd patent for the U.S.
The standard errors in Fig. \ref{fig:tech} look large at the bin of the 16th repetition.
This is because the bin is calculated for 9-16 repetitions.
Standard errors for every repetition is smaller than those in Fig. \ref{fig:tech}.
We can say that, to avoid decline, a team should incorporate a new technological field that they have no experience with where possible.

\begin{figure*}[h!]
  \begin{center}
    \includegraphics[scale=0.6, bb=0 0 500 500]{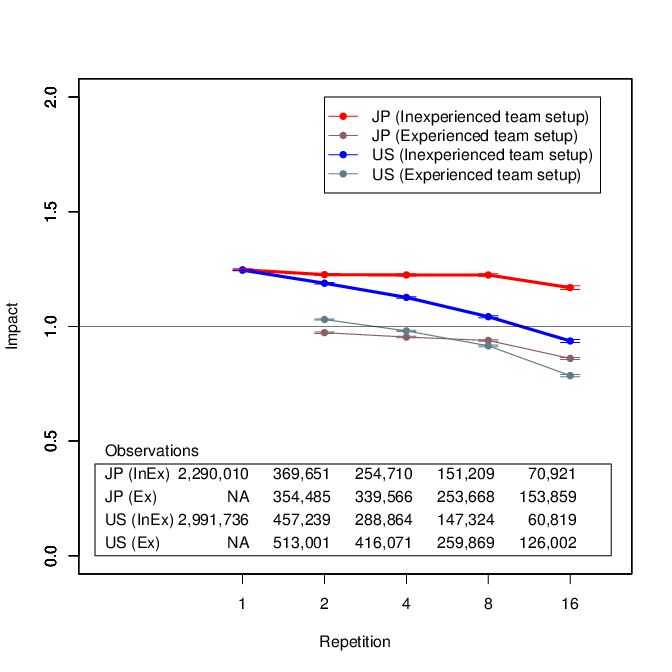}
  \end{center}
  \caption{Average impact of repeat collaborations categorized by development of team members.
Horizontal axis is repetitions by inventor pairs.
Vertical axis is average impact.
Brackets show standard errors.
``Inexperienced team setup'' (InEx) means some members other than target pairs enter or exit.
``Experienced team setup'' (Ex) means the opposite.
Horizontal axis is cut so that number of samples is greater than approximately 100.
Black horizontal line is impact $=1$. \label{fig:teamdiff}}
\end{figure*}

The second strategy is developing inventor diversity.
The definition of repeat collaborations thus far has been consecutive patents
that have been published by the same team.
We cannot count repeat collaborations with inventor diversity, that is to say, mixing teams,
by using the original definition
because if an inventor is different between two teams, these are considered as different teams.
Therefore, we need to redefine repeat collaborations to discuss this strategy.
We define repeat collaborations in another way on the basis of pairs of inventors.
Regardless of other inventors in collaborations,
repetitions are counted if two particular inventors are involved.
For example, if there is a series of collaborations with inventors, such as inventors A, B, and C, and inventors A, B, and D,
there are two repeat collaborations for inventors A and B but only one repeat collaboration for other pairs of inventors.
On the basis of the definition, we can consider the entrance and exit of other inventors in teams involving two specific inventors.
Fig. \ref{fig:teamdiff} plots the results.
There are two groups of results: inexperienced and experienced team setups.
If a patent involves two inventors
and they have not experienced a team setup [other teammate(s)] before,
the patent is categorized into an inexperienced team setup and vice versa.
The repetition number means the number of patents applied for by the same pair of inventors.
The plots include all patents by pairs regardless of whether patents are hits or not.
The results reveal that repeats with inexperienced team setups are better than those with experienced team setups.
Moreover, decline itself is highly mitigated in Japanese patent data.

We again conduct the Wilcoxon rank-sum tests between inexperienced and experienced team setups.
The inexperienced team setup is statistically dominant at the five percent level at every repetition in both the Japanese and U.S. data.
It must be noted again that there are not large enough sample sizes after a lot of repetitions.
We cannot stably see the significance after the 46th patent for Japan and the 30th patent for the U.S.
Therefore, we can say that changing a team setup where possible is always a good strategy to prevent degeneration.
Guimer\`a et al.'s seminal work \cite{Guimera05} showed repetitions negatively affect overall performance.
It can be said that our study adds new findings to their work in the sense that
changing a team setup causes a positive result at every step of repetition.

The two analyses, technological diversity and inventor diversity, cannot be compared simply
because their definitions of teams are different.
Even so, if we look at the gaps and the size of the error bars between the experienced and the inexperienced teams,
it would be plausible that mixing a team is more secure than learning a new technology with an existing team
because the relative size of the gaps in the standard errors of inventor diversity looks larger than those of technological diversity.

The dilemma of innovation comes from the contradiction
that we need to exploit our previously obtained resources, including shared knowledge and team coordination,
and simultaneously need to explore new solutions that have not been created yet.
From the viewpoint of exploitation,
it is easily imagined that teams will maximize the utility of their resources
and achieve inventions that gradually degenerate.
Therefore,
it is straightforward to give stimuli or perturbations to teams to expand their resources.
However, trying a whole new field of technologies and entirely destroying team coordination
eliminate the merit of exploiting.
Our two analyses done to show the effect of trying new technologies and new team setups
are evidence that there seem to be approaches to tackling the dilemma,
especially when we carefully look into the situation of knowledge sharing and team coordination.
In line with the discussion,
a preceding experimental study
indicated that mixed teams are more adaptive than intact teams
in the sense that the former explores the space of solutions without any intervention \cite{Gorman10}.
Also, another study revealed that
mixed teams are more promising than intact teams if retention intervals are long \cite{Gorman11}.
Studying the innovator's dilemma with massive data has only just begun,
and there is a lot of room to further discuss tackling the dilemma.

Furthermore,
as an implication of this study,
we discuss the applicability of our findings to other types of data.
We discussed repetitions of inventor groups and their dilemma in this study.
Since no matter what a group is, its aim is to achieve better performance,
we will probably observe the dilemma between routines and dysfunction
when a group does not adapt itself to situational changes.
There are preceding studies that discussed routines in groups in other contexts,
including political decision-making \cite{Janis82},
research and development projects \cite{Katz82},
or controlling uninhabited aerial vehicles \cite{Gorman10, Gorman11}.
Since we can probably define routines and dysfunctions in these groups,
we can expect that the dilemma, i.e. when or how to mitigate dysfunction, can be studied on the basis of the framework we proposed in this paper.

\section*{Conclusion}

We focused on repeat collaborations in creative projects
and how they affect performance.
Repeat collaborations have two contradictory aspects.
The first is a positive aspect: team development or experience.
The second is negative: team degeneration or decline.
This dilemma has not previously been quantitatively analyzed.
We provided extensive and quantitative analyses of the dilemma with creative projects
by using patent data from Japan and the United States.

We proposed three predictions to validate the dilemma.
Prediction 1 is that the greater the patent a team achieves, the longer the team will work together.
Prediction 2 is that the impact of consecutive patents decreases after a team makes hits.
Prediction 3 is that the expectation of impact by switching teams is greater than that from consecutively working in the same team
at some point during subsequent patents.
We found these predictions are validated on the basis of the data.

We then proposed preventive strategies against degeneration.
One is developing technological diversity, and another is developing inventor diversity.
We found the two strategies are both effective by validating them on the basis of the data.

We studied the relationship between the repetition of collaborations and performance.
However, retention between repetitions is not constant,
and the diversity of the retention can affect the performance.
A precedent study discussed this from the viewpoint of changing team memberships, which is the same approach as our study, 
in a controlled environment \cite{Gorman11}.
A complementary study with our data on the retention can be potential future work.







\section*{Acknowledgments}
This work was supported by KAKENHI 15K01217.
We thank Yang-Yu Liu for the valuable discussions.

\section*{{\small Author Contributions Statement}}
H. I conducted all analyses and wrote the manuscript.


%
%
%

\bibliography{paper}




\end{document}